\documentstyle[multicol,prb,aps,epsf]{revtex}
\begin{document}
\title{Optical conductivity in A$_3$C$_{60}$ (A=K, Rb)  }

\author  {J. van den Brink${}^{1,2}$, O. Gunnarsson${}^{1}$ 
and V. Eyert${}^{1,3}$}
\address{${}^{1}$Max-Planck-Institut f\"ur Festk\"orperforschung,
 D-70506 Stuttgart, Germany}
\address{${}^{2}$Laboratory of Solid State and Applied Physics, 
Materials Science Centre, University of Groningen, Nijenborgh 4, 
9747 AG Groningen, The Netherlands}
\address{${}^{3}$Hahn-Meitner-Institut, Glienicker Stra\ss e 100,
D-14109 Berlin, Germany}

 \date{\today}

\maketitle

\begin{abstract}
We study the optical conductivity in A$_3$C$_{60}$ (A =K, Rb). The effects
of the electron-phonon interaction is included to lowest order in the 
coupling strength $\lambda$. It is shown that this leads to a narrowing of the
Drude peak by a factor $1+\lambda$ and a transfer of weight to a mid-infrared
peak at somewhat larger energies than the phonon energy. Although this goes
in the right direction, it is not sufficient to describe experiment.
\end{abstract}

\begin{multicols}{2}
\section{Introduction}
The optical conductivity in A$_3$C$_{60}$ (A=K, Rb) has an unusual 
and interesting behaviour.\cite{Iwasa,Degiorgi92,Degiorgi94,Science,Com}
The weight of the Drude peak is reduced
by one order of magnitude relative to the weight    for free electrons 
with the appropriate band mass. 
Much of the missing weight appears instead in a ``mid-infrared'' structure
at about 0.06 eV. This suggests very strong interaction effects, e.g.,
electron-phonon or Coulomb interaction. The understanding of the
optical absorption could therefore contribute much to the understanding 
also of other properties of A$_3$C$_{60}$.

A$_3$C$_{60}$ has orientational disorder, with the C$_{60}$ molecules      
taking, more or less randomly, one out of two preferential 
orientations.\cite{Stephens}
This orientational disorder leads to a substantial modification of the 
optical conductivity in one-particle calculations.
 For an ordered system, the Drude peak collapses to a    
$\delta$-function, while the 
disorder leads to a broad Drude peak.\cite{Gelfand}
The calculated optical conductivity furthermore shows a structure at
somewhat larger energies than the experimental mid-infrared 
structure,\cite{Gelfand} although the structure is less pronounced 
and at higher energy than
in the experimental spectrum. More serious is, however, that the weight 
and width of the Drude peak 
are much larger than the experimental results. Although it is hard to 
separate the theoretical results in a Drude and a mid-infrared
structure, it may be estimated that the theoretical Drude width  is 
more than a factor of ten too large. 

The strong reduction of the Drude width  suggests very strong 
renormalization effects, e.g., due to the electron-phonon or
electron-electron interactions. The fullerenes have phonons
with an energy of about 0.06 eV, which show a strong coupling 
to the electrons.\cite{C60RevMod} Since these phonons may transfer
weight from the Drude peak to a mid-infrared structure, we here study 
the effect of phonons.  

We limit ourselves to calculating the electron self-energy to lowest
order. This is       sufficient if Migdal's theorem\cite{Migdal} is valid. 
It is,
however, questionable if this is true for the fullerenes, and we should
keep in mind that higher order effects may be important.
To obtain the optical conductivity we calculate the current-current
response function. We can neglect vertex corrections,\cite{vertex}
 since the electron self-energy is ${\bf q}$-independent in our approach.
The current-current response function is then reduced to a product
of two electron Green's function.
We find that the electron-phonon coupling leads to a narrowing of the 
Drude peak by about a factor of $1/(1+\lambda)$, where $\lambda$
is the is the electron-phonon coupling constant. Although this goes in 
the right direction,
it is by far not sufficient to explain the experimental data. 

In Sec. II we present the formalism and the model. In Sec. III we show the 
results and in Sec. IV multiplet effects are briefly discussed. 
The results and other possible explanations of the narrow Drude
peak are discussed in Sec. V.

\section{Formalism and model}
The optical conductivity is given by\cite{Mahan}
\begin{equation}\label{eq:1}
{\rm Re} \sigma_{\alpha \beta}(\omega)={\rm Re} {{\rm lim}\atop {\bf q}\to 0}
{i\over \omega} \pi_{\alpha \beta} ({\bf q},\omega)
\end{equation}
where 
\begin{equation}\label{eq:2}
\pi_{\alpha \beta}({\bf q},\omega)=-i\int_0^{\infty}dt e^{i\omega t}
\langle 0 |\lbrack j_{\alpha}^{\dagger}({\bf q}, t),j_{\beta}({\bf q},0)
\rbrack | 0\rangle.
\end{equation}
Here $j$ is the current operator and $|0\rangle$ is the ground-state.
Below, we use a formalism where the electron self-energy is 
${\bf q}$-independent. It can then be shown\cite{vertex} that 
the vertex corrections in the current-current response function vanish 
for ${\bf q}\to 0$, due to the odd parity of the current operator. 
We can then write the optical conductivity as a product of two 
Green's functions, only keeping a simple bubble of dressed Green's
functions in the diagrammatic expansion of $\sigma$.
If we express the current operator as
\begin{equation}\label{eq:3}
j_{\alpha}=\sum_{\sigma}\sum_{nn^{'}}v^{\alpha}_{nn^{'}}c_{n \sigma}^{\dagger}
c_{n^{'}\sigma},
\end{equation}
the optical conductivity is given by
\begin{eqnarray}\label{eq:4}
{\rm Re} \ \sigma_{\alpha \beta}&&={2\over \omega}{\rm Re} \sum_{nn^{'}}
\sum_{mm^{'}}v_{nn^{'}}^{\alpha \ast}v_{mm^{'}}^{\beta}  \nonumber  \\
&&\times
\int_{-\infty}^{\infty} {d\omega^{'}\over 2 \pi}G_{nm^{'}}(\omega+\omega^{'})
G_{mn^{'}}(\omega^{'})
\end{eqnarray}
where $G$ is the electron Green's function.
This can be rewritten as\cite{Schweitzer}
\begin{eqnarray}\label{eq:4a}
&&{\rm Re} \ \sigma_{\alpha \beta}={2\pi\over \omega} \sum_{nn^{'}}
\sum_{mm^{'}}v_{nn^{'}}^{\alpha \ast}v_{mm^{'}}^{\beta}    \\
&&\times
\int_{-\infty}^{\infty} d\omega^{'}A_{nm^{'}}(\omega+\omega^{'})
A_{mn^{'}}(\omega^{'})\lbrack f(\omega^{'})-f(\omega^{'}+\omega)\rbrack,
\end{eqnarray}
where  $A_{nm}(\omega)={\rm Im} \ G_{nm}(\omega-i0^{+})/\pi$ and $f(\omega)$
is the Fermi function.

We consider the three $t_{1u}$ orbitals of C$_{60}$ which are 
connected by hopping matrix elements $t$
\begin{equation}\label{eq:4b}
H^{\rm el}=\sum_{i\sigma}\sum_{m=1}^3\varepsilon_{t_{1u}}n_{im\sigma}+
\sum_{<ij>\sigma
mm'}
t_{ijmm'}\psi^{\dagger}_{im\sigma} \psi_ {jm'\sigma}
\end{equation}
The orientational disorder\cite{Stephens} has been built into the
matrix elements $t_{ijmm'}$.\cite{Orientation,Satpathy,MazinAF}
Deshpande {\it et al.} have used a similar model for calculating the     
phonon self-energy.\cite{Deshpande} 
We want to describe the coupling to the intramolecular five-fold 
degenerate $H_g$ Jahn-Teller modes. Due to the intramolecular character, the 
coupling has a local form.
To describe the electron-phonon interaction, we use the Hamiltonian
\begin{eqnarray}\label{eq:4d}
H^{\rm el-ph}&&=
 \omega_{ph}\sum_{m=1}^5(b^{\dagger}_{ m}b_{ m}+{1\over 2})\nonumber \\ 
&&+ {g\over 2} \sum_{m=1}^5 \sum_{\sigma}\sum_{i=1}^3
\sum_{j=1}^3
V_{ij}^{(m)}\psi^{\dagger}_{i\sigma} \psi_{j\sigma}(b_{ m}+ b_{ m}^{\dagger}),
\\  
\end{eqnarray}
where $\omega_{ph}$ is the a phonon frequency, $b_m$ annihilates
a phonon with quantum number $m$,
$V^{(m)}_{ij}$ are dimensionless coupling constants\cite{Lannoo,c60jt} 
given by symmetry and $g$ is an overall coupling strength. 
The electron-phonon coupling constant $\lambda$ is then given by
\begin{equation}\label{eq:4c}
\lambda={5\over 3}N(0){g^2\over \omega_{ph}},
\end{equation}
where $N(0)$ is the density of states per spin at the Fermi energy.

We now construct a consistent current operator, essentially 
following Ref. \onlinecite{czycholl}. 
We write the density $\rho(i)$ at a site $i$ as
\begin{equation}\label{eq:6}
\rho(i)=\sum_{m\sigma}\psi^{\dagger}_{im\sigma}\psi_{im\sigma}.
\end{equation}
Here we only consider the number of electrons on a given site, and neglect
the possible polarization of the charge on this C$_{60}$ molecule. 
Due to this assumption
we obtain no terms in the current operator describing on-site transitions.
Since the transitions between $t_{1u}$ orbitals on the same site are forbidden,
Eq. (\ref{eq:6}) is sufficient for our purposes. 
Imposing charge and current conservation
\begin{equation}\label{eq:7}
{\bf q}\cdot {\bf j(q)}=-e [H,\rho({\bf q})],
\end{equation}
we obtain 
\begin{equation}\label{eq:8}
{\bf q}\cdot {\bf j(q)}= - {ie\over \sqrt{N}}\sum_{ijmm^{'}}t_{ijmm^{'}}
{\bf q}\cdot ({\bf R}_i-{\bf R}_j) 
\end{equation}
in the limit ${\bf q}\to 0$. Here ${\bf R}_i$ is the position of 
molecule $i$. We then obtain the current matrix elements
\begin{equation}\label{eq:9}
v_{im,jm^{'}}^{\alpha}=-iet_{ijmm^{'}}(R_i^{\alpha}-R_j^{\alpha}).
\end{equation}

The electron self-energy is calculated to lowest order in the
electron-phonon interaction. 
\begin{equation}\label{eq:10}
\Sigma_{nn^{'}}^{\rm El-phon}(\omega)=i\sum_{m \mu}\int {d \omega^{'} \over 2 \pi} 
\lambda^{\mu}_{nm} G_{mm}^{(0)}(\omega -\omega^{'})D^{(0)}_{\mu \mu}   
(\omega^{'})\lambda^{\mu}_{mn^{'}},
\end{equation}
where $G^{(0)}_{mm}$ and $D^{(0)}_{\mu \mu}$ are the zero order electron and
phonon Green's functions, respectively.     
The electron-phonon coupling is described by $\lambda^{\mu}_{nm}$,
which is expressed in terms of the coupling constants $V^{(m)}_{ij}$ 
and the one-particle solutions.
 The interacting electron Green's function 
is then obtained from the Dyson's equation
\begin{equation}\label{eq:11}
G(\omega)=G^{(0)}(\omega)+G^{(0)}(\omega)\Sigma(\omega)G(\omega),
\end{equation}
where a matrix notation has been used.

We next discuss qualitatively how the optical conductivity may
change due to the electron-phonon interaction. If the band width is much 
larger than a typical phonon frequency, Migdal's theorem\cite{Migdal}
is valid. For states with an energy smaller than the phonon energy,
the quasiparticle energy is then reduced by a factor\cite{Grimvall} 
\begin{equation}\label{eq:12}
Z={1 \over 1+\lambda},
\end{equation}
where $\lambda$ is the electron-phonon coupling. Furthermore, the 
quasiparticle weight is reduced by the same factor.\cite{Grimvall}
For A$_3$C$_{60}$ it is very questionable if Migdal's theorem is valid,
and interesting effects happen due to the fact that the band width
is not much larger than the phonon frequencies.\cite{Liechtenstein}
Nevertheless, we can expect to obtain some insight into the effect of the
electron-phonon interaction by making the above assumptions,
i.e., assuming that the electrons can be treated as noninteracting 
but with weights and energies which are reduced by a factor $Z$.
For $\omega>0$ we then have
\begin{equation}\label{eq:13}
\sigma_{\alpha \alpha}(\omega)\sim {{\rm lim} \atop {\bf q} \to 0} 
{1\over \omega}
{\rm Im } \sum_{n}^{unocc}\sum_m^{occ}
 { |\langle n|j_{\alpha}({\bf q}) |m \rangle|^2
\over \omega -\varepsilon_n +\varepsilon_m-i0^{+}}
\end{equation}
We replace $\varepsilon_n$ by $Z\varepsilon_n^{(0)}$ and 
$\langle n|j_{\alpha}({\bf q}) |m \rangle$ by 
$Z\langle n|j_{\alpha}({\bf q}) |m \rangle^{(0)}$, where the suffix 
$0$ refers to the noninteracting system. This leads to
\begin{equation}\label{eq:14}
\sigma_{\alpha\alpha}(\omega)=\sigma^{(0)}_{\alpha \alpha}({\omega \over Z}),
\end{equation}
where $\sigma^{(0)}$ is the optical conductivity without the 
electron-phonon interaction. For zero frequency $\sigma$ is unchanged,
as it should, since the resistivity $\sigma(0)$ is not influenced by the
electron-phonon interaction at zero temperature, considered here.
We can see, however, that the energy scale is reduced by a factor
of $Z$, and that the weight of the Drude peak is reduced correspondingly. 
For larger frequencies these considerations are of course too simple,
since we then have to consider the whole Green's function including 
phonon satellites and not just the quasiparticle.

\noindent
\begin{figure}[bt]
\unitlength1cm
\begin{minipage}[t]{8.5cm}
\centerline{\epsfxsize=3.375in \epsffile{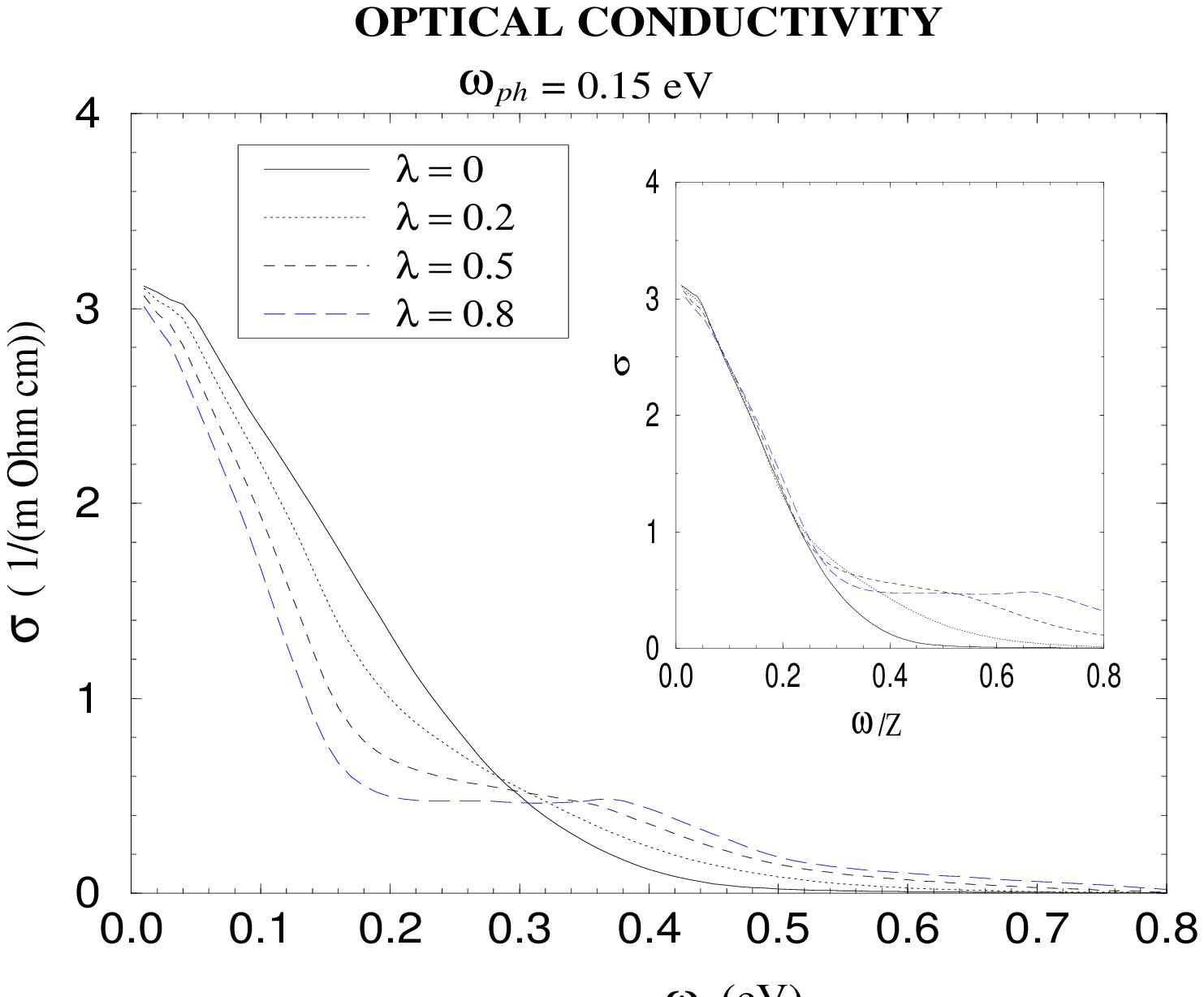}}
\vskip0.3cm
\caption[]{\label{fig1}
Optical conductivity $\sigma(\omega)$ for the phonon frequency
$\omega_{ph}=0.15$ eV and for different electron-phonon coupling
constants $\lambda$. The figure illustrates how the Drude peak becomes
narrower and how weight is transferred to a mid-infrared peak as
$\lambda$ is increased. The inset shows $\sigma$ as a function of 
$\omega/Z$, where $Z=1+\lambda$. This illustrates how the width of the
Drude peak is reduced by a factor of $1+\lambda$ due to the 
electron-phonon interaction.}
\end{minipage}
\hfill
\end{figure}

\section{Results}

\begin{figure}[bt]
\unitlength1cm
\begin{minipage}[t]{8.5cm}
\centerline{\epsfxsize=3.375in \epsffile{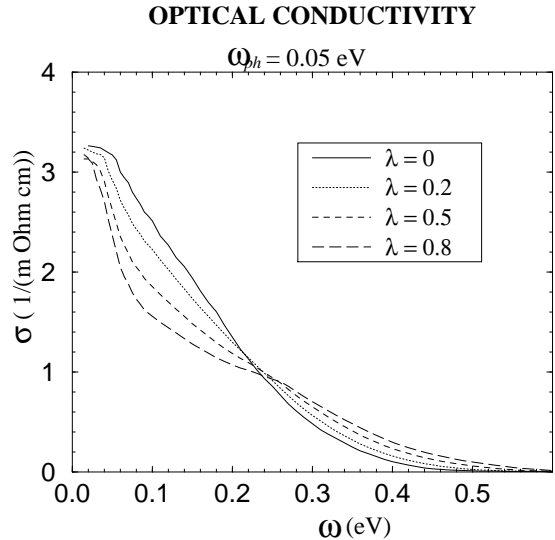}}
\caption[]{\label{fig2}
The same as in Fig. \ref{fig1} but as a function of $\omega/Z$,
where $Z=1/(1+\lambda)$. The similarity of the curves for small
values of $\omega/ Z$ but different values of $\lambda$ illustrates
how the Drude peak width is reduced by a factor $Z$.  }
\end{minipage}
\hfill
\end{figure}

In Fig. \ref{fig1} we show the optical conductivity for a phonon
frequency $\omega_{ph}=0.15$ eV.
Without electron-phonon coupling ($\lambda=0$) the spectrum shows a
broad Drude peak. As $\lambda$ is increased, the Drude peak becomes
narrower and weight is transferred to a structure in the energy range
0.2-0.4 eV. In the inset in Fig. \ref{fig1} the same results are shown 
as a function of $\omega/Z$. 
The curves now essentially fall on top of each other for small $\omega$.
This illustrates the result (\ref{eq:14}) that the width of the Drude peak 
is reduced    by a factor $Z$.
Fig. \ref{fig2} shows the results for a lower   
phonon frequency $\omega_{ph}=0.05$ eV. The spectrum is similar as in
Fig. \ref{fig1}, but the mid-infrared structure has moved to lower   
frequencies.

\noindent

From photoemission for a free C$_{60}^{-}$ molecule\cite{c60m}
 and from neutron scattering\cite{Prassides}
 it has been estimated that the strongest coupling 
is to the second lowest H$_g$ mode at about 0.054 eV.
From Raman scattering the strongest coupling was found for the
lowest mode at about $\omega_{ph}=0.033$ eV.\cite{Kuzmany}
The value $\omega_{ph}=0.05$ used in Fig. \ref{fig2} should therefore
be more realistic then the one in Fig. \ref{fig1}, and one might even
argue for a still smaller value of $\omega_{ph}$. This would then tend
to give an energy of the mid-infrared structure of the right order of
magnitude, although it is still larger than the experimentally 
observed value   0.06 eV.            
The electron-phonon coupling is of the order 
$\lambda\sim 0.5-1.0$.\cite{C60RevMod} The width of the Drude peak is
then reduced by a factor of 1.5-2.
This reduction goes in the right direction, but it is much too small
to explain experiment.

\section{Multiplet effects}

An alternative mechanism for transferring weight from the Drude peak to 
the mid-infrared peak is provided by multiplet effects. Within the
$t_{1u}$ system, these are described by the exchange integral $K$
between two $t_{1u}$ orbitals and the difference  
$\delta U\equiv U_{xx}-U_{xy}$ between the direct
Coulomb integral for equal and unequal orbitals. Here we use 
$\delta U=2K$. The C$_{60}^{3-}$ molecule has a ground-state 
with spin 3/2 and states with the spin 1/2 at $3K$ and $5K$ above
the ground-state.  The value of $K$ has been estimated to be
0.05 eV,\cite{Martin} and 0.024 eV.\cite{Joubert}  
The unscreened value has been found to be $K=0.12$ eV and within RPA
screening $K=0.030$ eV.\cite{Ferdi}
The experience from atomic multiplets is that these are only weakly
reduced ($\sim 20 \%$) relatively to what is predicted by the 
unscreened Coulomb integrals, both for free atoms and for 
solids.\cite{Cowan} We also find that to describe the multiplets
in the $h_u-t_{1u}$ exciton, unscreened integrals give a splitting
of the right order of magnitude. Due to the lack of extensive experience for 
the large C$_{60}$ molecule, we nevertheless consider the whole 
range of estimates for the multiplet integrals below.
If the lower values of these estimates are used,
the multiplet splitting is of the same order
of magnitude as the energy of the mid-infrared structure, and it is
 then interesting to study to what extent these effects can 
explain this structure.

We have added a multiplet interaction to the   
Hamiltonian in Eq.~(\ref{eq:4b})
\begin{eqnarray}\label{eq:m1}
H_U=&&{2 \over 3}\delta U\sum_{im} n_{im\uparrow}n_{im\downarrow}
-{1\over 3}\delta U\sum_{i\sigma\sigma^{'}}\sum_{m<   m^{'}}n_{i\sigma m}
n_{i\sigma^{'}m^{'}}  \nonumber  \\
+&&{1\over 2}K \sum_{i\sigma\sigma^{'}}\sum_{m\ne m^{'}}
\psi^{\dagger}_{i\sigma m}\psi^{\dagger}_{i\sigma^{'} m^{'}}
\psi_{i\sigma^{'} m}\psi_{i\sigma m^{'}}   \\
+&&{1\over 2}K \sum_{i\sigma}\sum_{m\ne m^{'}}
\psi^{\dagger}_{i\sigma m}\psi^{\dagger}_{i-\sigma m}
\psi_{i-\sigma m^{'}}\psi_{i\sigma m^{'}}.  \nonumber  
\end{eqnarray}
The simple Coulomb interaction
\begin{equation}\label{eq:m1a}
H_{U}^0=
U\sum_i\sum_{(\sigma m) < (\sigma'm')}n_{i\sigma m}n_{i\sigma'm'},
\end{equation}
should also be added but is not considered here, since in simple treatments
it does not give a contribution to the mid-infrared structure.

We have estimated the self-energy to second order in $\delta U$ and $K$
and obtained                 
\begin{equation}\label{eq:m3} 
\Sigma_{nn\sigma}^{\rm Mult} \sim {K^2\over W}.
\end{equation}
 This has to be compared 
with the self-energy due to the electron-phonon energy, which is
of the order 
\begin{equation}\label{eq:m4}
\Sigma_{nn\sigma}^{\rm El-phon} \sim \lambda \omega_{ph}.
\end{equation}
If we put $K=0.03$ eV, $W=0.5$ eV, $\lambda=1$ and $\omega_{ph}=0.1$ eV,
we find that $\Sigma^{\rm El-phon}$ is about a factor of 35 larger than
$\Sigma^{\rm Mult}$.
This suggests that although the multiplet effects may transfer
weight to the mid-infrared peak, the effect should be very small. 
If, on the other hand, we use a large value $K=0.15$ eV for the 
multiplet integral, the second order self-energy due to the multiplet
integrals becomes comparable to the 
electron-phonon contribution. In this case, however, the multiplet splitting
is much larger than the energy of the mid-infrared peak. It therefore
seems likely that the multiplet effects treated in second order theory
cannot explain the energy and weight of the mid-infrared peak. 
We observe, however, the second order perturbation theory used here
is not sufficient to describe the atomic limit, and that a better
treatment conceivably could change the conclusions somewhat.
  
\section{Discussion}
We have calculated the optical conductivity, including the effects 
of the lowest order self-energy diagram due to the electron-phonon
interaction. This coupling reduces the width of the Drude peak and transfers
weight to the mid-infrared structure at an energy somewhat larger  
than the phonon
frequency. This leads to a mid-infrared structure with an energy of
the right order of magnitude, but a bit too large. 
The reduction of the width of the Drude peak goes in the right direction,
but it is much too small. We observe that the self-energy was calculated 
under the assumption that Migdal's theorem is valid. Since Migdal's theorem
is questionable for these systems, higher order corrections could modify
these conclusions. 

It is interesting that Liechtenstein {\it et al.}\cite{Sasha} found
a rather narrow Drude peak (width $\sim$ a few hundredths of an eV) 
in a one-particle calculation. 
As mentioned before, the C$_{60}$ molecules in A$_3$C$_{60}$ have primarily 
two different orientations. It has been found on theoretical grounds that 
it is energetically favorable if neighboring C$_{60}$  molecules have 
different (``antiferromagnetic'') orientations. \cite{Orientation,MazinAF}
The system can then be mapped onto a frustrated Ising model, 
for which the ground-state has a frustrated  antiferromagnetic 
ordering.\cite{MazinAF} This ordering  leads to the narrowing of the Drude 
peak mentioned above.\cite{Sasha}
Experimentally, a tendency to a short-ranged ``antiferromagnetic''
correlation has been found,\cite{Teslic} but under normal experimental
conditions the samples are apparently cooled too fast to develop the 
long-ranged partial order assumed in Ref. \onlinecite{Sasha}.
It therefore does not seem likely that the partial ordering assumed 
in Ref. \onlinecite{Sasha} explains the narrow Drude peak in experimental
samples used so far.

It is interesting to ask what other effects may contribute to the explanation
of the optical conductivity. We have illustrated that multiplet effects
are unlikely to explain the experimental results, at least if they are 
treated to lowest order. These systems have a strong coupling to a
charge carrier plasmon at 0.5 eV due to the oscillations of the three
$t_{1u}$ electrons.\cite{Sohmen,Eyert,Liechtenstein} 
In analogy with the coupling to the phonons, one may argue that the
plasmons have a coupling constant $\lambda_{pl}\sim 2.5$.\cite{gwc60}
 Taking over 
the arguments from the electron-phonon coupling one might then expect
a substantial narrowing from the coupling to the plasmons.
This picture is, however, too simple, and a calculation of the electron
self-energy in the so-called GW approximation\cite{Hedin} shows only a
modest reduction of the band width.\cite{gwc60} Actually, estimates of
the specific heat\cite{Ramirez,Meingast} do not show an enhancement 
compared with the result obtained from band structure calculations, 
apart from the enhancement expected from an electron-phonon interaction
with a $\lambda\sim 0.5-1$.
If these estimates are correct, they suggest that
many-body interactions do not reduce the dispersion in A$_3$C$_{60}$
(A= K, Rb). This is also consistent with the
susceptibility,\cite{Ramirez} which shows a very weak temperature
dependence, implying that there is no narrow peak in the density
of states. We should then not expect an explanation 
of the narrow Drude peak in terms of a mechanism which reduces the dispersion
beyond the reduction due to the electron-phonon interaction.
Instead we should search for a mechanism which influences a two-particle
spectrum, like the optical conductivity, without increasing the 
effective mass.  

\section{Acknowledgements} 
JvdB thanks the Stichting Scheikundig Onderzoek Nederland for financial
support.

\end{multicols}
\end{document}